# Periodic oscillations of flux flow resistance in intrinsic Josephson junctions with lateral sizes down to a size of non-linear Josephson vortex core

S. Urayama[1,2], T. Hatano[1,2], H. B. Wang[1], M. Nagao[1], S. M. Kim[1], and J. Arai[2]

[1] National institute for Materials Science (NIMS), Tsukuba 305-0047, Japan,
[2]Tokyo University of Science (TUS), Yamazaki, Noda, Chiba 278-8510, Japan

*Abstract*— To investigate in-phase structures (rectangular lattice) of Josephson vortex lattice (JVL) in intrinsic Josephson junctions (IJJs) of $Bi_2Sr_2CaCu_2O_{8+\delta}$, we fabricated IJJs with various lateral sizes down to a size of non-linear core of Josephson vortex using focused-ion-beam etching and measured flux flow resistance ($R_{FF}$) with in-plane magnetic fields ($H$). We have found that above $H_1 \equiv \frac{\Phi_0}{\gamma \cdot s^2}$, a size-independent critical field, the critical current shows Fraunhofer pattern in narrow-length IJJs (0.29-2.77 μm), and then the $R_{FF}$~$H$ oscillates with $H_0$ ( $\equiv \frac{\Phi_0}{L \cdot s}$ ) period. Most strikingly, such oscillations become pulse-like in sub-micron IJJs, indicating that between the pulses, the rectangular JVLs become major structures.

## I. INTRODUCTION

It is well known that alternating superconducting layers and insulating layers in high-$T_c$ superconductors work as an array of intrinsic Josephson junctions (IJJs) [1,2]. The IJJs are expected to be a high-frequency oscillator based on the Josephson plasma excitation, of which frequencies can be 100 GHz to a few THz [3].

To generate THz emission, Josephson plasma excitation using flux (Josephson vortex (JV), generated by magnetic fields ($H$) parallel to an *ab*-plane of IJJs) flow is proposed [4]. It is said that a crucial requirement for the THz emission is coherent in-phase oscillations of all layers in IJJs. In the case of the Josephson plasma excitation using the flux flow, the formation of rectangular JVs lattice (JVL) is necessary. However, it is known that vortices generally form triangular lattice because of the repulsive interaction among them. Therefore, the control of the vortex lattice structure is an important issue for the THz emission.

The periodic oscillations of flux flow resistance ($R_{FF}$) against $H$ ($R_{FF}$ -$H$) under the *c*-axis bias currents, reported by Ooi *et al.*, gives us a chance to know the lattice structure [5]. The period corresponds to the field needed to add “one” JV per “two” IJJs, which is expressed as $H_0$ /2 ( $\equiv \frac{\Phi_0}{2L \cdot s}$ , where $L$, $\Phi_0$, and $s$ are the sample width, the flux quantum, and the layer periodicity 1.54 nm, respectively). The oscillations can be explained by the formation of triangular JVLs. At high fields and high bias currents, the period was found to be doubled, which is expressed as $H_0$ ( $\equiv \frac{\Phi_0}{L \cdot s}$ ). It was said that the doubled period originates from the formation of rectangular JVLs [6]. We name the former oscillations “$H_0$ /2 oscillations”, and the latter “$H_0$ oscillations”.

Subsequently, the $H_0$ /2 oscillations have been reproduced numerically with taking the surface barrier effect into account for the entry and escape of JVs at the edges [7, 8]. Koshelev described schematically an *H*-*L* phase diagram of JVL and introduced lattice deformation length $l_b$ that gives a measure of edge effects. For finite-sized IJJs Machida has predicted the alternating appearance of triangular JVLs ( $H \approx n \cdot H_0$ ) and rectangular JVLs ( $H \approx (n+1/2) \cdot H_0$ ) with increase of $H$, in other words, triangular JVLs also exist at the fields where $H_0$ oscillations are observed [9,10].

Experimentally, Kakeya *et al.* investigated the $L$-dependence systematically with the junction length from 1.8 μm to 9.4 μm [11]. They found a linear relation between $L$ and a threshold field $H_{th}$ where the periodicity changes from $H_0$ /2 to $H_0$. Hence they suggested that the rectangular JVLs are favorable for smaller junctions. Hatano *et al.* studied finite-sized IJJs (L=1.8 μm) to enhance the edge pinning of the JVL and discussed the mechanism of oscillation in $R_{FF}$ and the JVL structures [12].

In addition to the above-mentioned results, much stronger edge effect has been expected in sub-micron IJJs, thus the rectangular JVLs become stable, that is, in-phase oscillations of all layers can be expected. However, until now, $R_{FF}$ oscillations in sub-micron IJJs have not been studied yet.

In this paper, IJJs are fabricated with lengths from 2.77 μm to 0.29μm. The smallest one is comparable to the size of non-linear core of the JV ( $\gamma s$ , typically 0.3 μm in our samples where $\gamma$ is the anisotropy factor). We have found that above $H_1 \equiv \frac{\Phi_0}{\gamma \cdot s^2}$ , a size-independent critical field, the critical current shows Fraunhofer pattern in narrow-length IJJs (0.29-2.77 μm), and then $H_0$ oscillations appear. In addition, pulse-like periodic oscillations of $R_{FF}$ observed in our experiments imply that the rectangular JVLs become major structures between the pulses. Therefore it is of the great essence to employ narrow intrinsic Josephson junctions for practical applications.

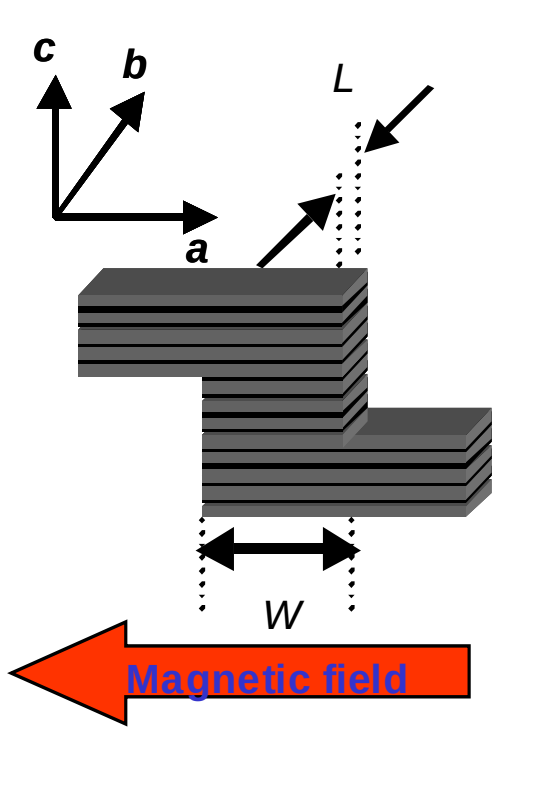

| Schematic view of IJJs | $L$/μm | $W$/μm | $T_c$/K |
|---|---|---|---|
| | 0.29 | 14.95 | 79 |
| | 0.38 | 15.05 | 78 |
| | 0.55 | 15.27 | 78 |
| | 0.59 | 15.07 | 77 |
| | 0.92 | 15.20 | 79 |
| | 1.15 | 15.12 | 79 |
| | 1.19 | 15.33 | 78 |
| | 1.31 | 14.36 | 77 |
| | 1.46 | 15.25 | 77 |
| | 1.74 | 15.09 | 80 |
| | 1.86 | 14.84 | 80 |
| | 2.32 | 15.36 | 78 |
| | 2.77 | 14.72 | 80 |

Table 1: Schematic view, dimensions of the length L (mm) and width W (mm), and Tc (K) of Bi-2212 IJJs studied in this paper.

## II. Experimental

$Bi_2Sr_2CaCu_2O_{8+\delta}$ (Bi2212) single crystal whiskers were grown by a Te-doping method [13]. After a whisker was fixed on an MgO substrate, electrodes were formed with silver paste. Subsequently, in-line type IJJs were fabricated using focused ion beam etching [14]. In our experiments, the lengths of junctions $L$ ($\perp H$) were systematically varied from 0.29 to 2.77 μm. For comparison, the width $W$ (//H) and the junction number of all samples were fixed at about 15 μm and 100, respectively. Detailed parameters and a schematic view of IJJs are shown in Table 1.

Electric transport properties were measured with a standard four-probe method using a Physical Property Measurement System (PPMS, Quantum Design), which can apply magnetic fields up to 7 T.

Prior to the measurements, the *ab*-plane alignment was precisely adjusted by angular dependence of the $R_{FF}$ under the constant magnetic field at 60 K. When the field direction is closely parallel to the *ab*-plane, the JVs will be in a lock-in state [15], which is free from pancake vortex. After the angular adjustment by setting the angle at the maximum of $R_{FF}$, we increased the temperature to 100 K, which is well above $T_c$, then decreased the temperature to a desirable one, and measured $R_{FF}$ by sweeping $H$ from 0 to 7 T with a constant *c*-axis current.

## III. Results and Discussion

We started our measurements from estimating the anisotropy factor γ. According to Koshelev, JVs start to form dense lattices at a starting field $H_{start} = \frac{\Phi_0}{2\pi\gamma s^2}$ [8], and then $H_0/2$ oscillations are expected to appear. Among several of our largest samples, $H_0/2$ oscillations start to appear at around 0.7 T (not shown here because of small amplitude). Accordingly, γ is estimated to be ~200. The product $\gamma s$ , namely the size of non-linear Josephson vortex core, is about 0.3 μm in our samples.

The *L*-dependence of $R_{FF}$ (0.34~2.77μm) at 60 K with a bias current of 30 A/cm$^2$ is shown in Fig. 1, where (a) is plotted against $H$, and (b) normalized fields ( $H/H_0$ ). In Fig. 1(b), a grid interval line of 0.5 $H/H_0$ is drawn for clarity. As presented in Fig. 1(b), for longer IJJs, $H_0/2$ oscillations are observed at low $H$, and $H_0$ oscillations are observed at high $H$. For instance, in the case of $L$=2.32 μm, $H_0$/2 oscillations are observed below $5H_0$ (~2.8 T), and $H_0$ oscillations are observed above $7H_0$ (~ 4 T). Similar characteristics have been observed earlier [6, 11]. In contrast, for short IJJs only $H_0$ oscillations are observed in the case of $L$=0.38 μm which is comparable to $\gamma s$, as shown in Fig. 1.

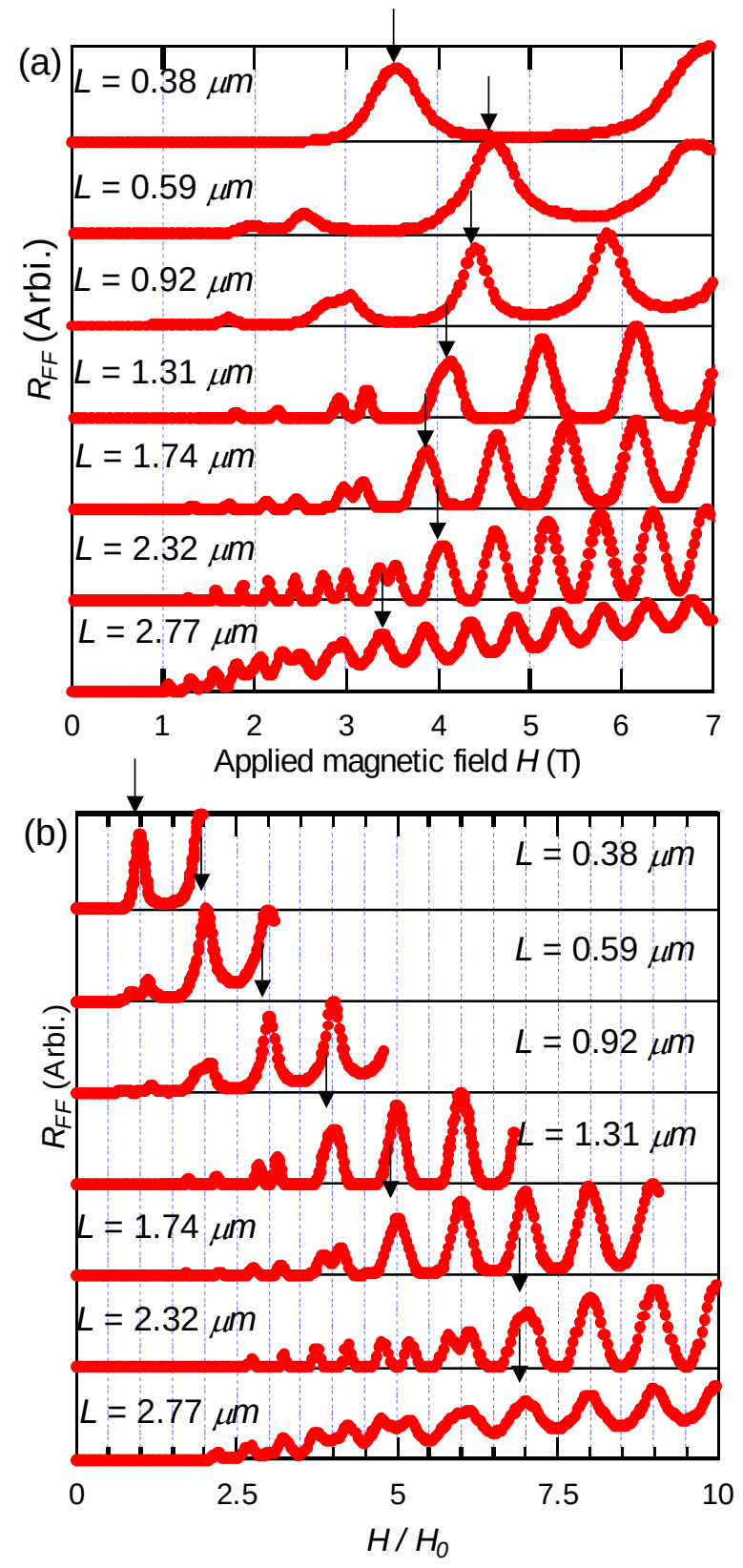

Fig. 1: Observed periodic oscillations of flux flow resistance (RFF) in Bi-2212 IJJs of 0.29 mm <L<2.77 mm with the c-axis current density 30 A/cm2 at 60 K; (a) plotted against applied magnetic fields H and (b) plotted against normalized fields H/H0. The arrows show the peaks above which H0 oscillations start.

As we know, in a short ($L<\lambda_J$, where $\lambda_J$ is Josephson penetration depth) single Josephson junction (SJJ), such as a conventional Nb/$AlO_x$/Nb Josephson junction, the critical current $I_c$ has a typical magnetic field dependence known as Fraunhofer pattern (the periodicity of the oscillations is $H_{0,\,\mathrm{SJJ}} \equiv \frac{\Phi_0}{L\cdot d}$, where $d$ is an effective barrier thickness). In our experiments for IJJs, we found that the peaks or valleys in the $H_0$ oscillations appear at the valleys or peaks in the Fraunhofer pattern as shown in Fig. 2. In other words, when the critical current has its local maximum at (n+1/2)$H_0$, the Josephson vortex flow is strongly suppressed. The fact that the Fraunhofer dependence can be explained by SJJ-like behavior in IJJs (intrra layer interaction of JVs is dominant compared to the interlayer interaction). This implies that the JVL is pinned as rectangular JVL at $H$=(n+1/2)$H_0$ where the $R_{FF}$ shows valley (rectangular JVLs form when the interaction along $c$-axis is not important.). At $H=nH_0$, where the edge pinning effect is the smallest, JVLs can flow easily. However, there is no evidence that the JVLs flow as rectangular JVLs. Therefore a safe conclusion can be drawn that the rectangular JVLs are stable at least in a static case and at $H$~(n+1/2)$H_0$.

We found that there exists a size-independent transition field $H_{tr}$ marked by arrows in Fig. 1, which is defined as the field where $n\cdot H_0$ starts to show peaks (Before $H_{tr}$, both $n\cdot H_0$ and $(n+1/2)\cdot H_0$ correspond to valleys in $H_0$/2 oscillations.). We use this definition because Fraunhofer pattern is well established above $H_{tr}$, which will be explained in the following paragraph. Noticeably, $H_{tr}$ is independent of $L$, about $4.2\pm0.5$ T, as shown in Fig. 3 (a).

For the appearance of $H_0$ oscillations, we consider that the following condition should be satisfied in IJJs. Non-linear core

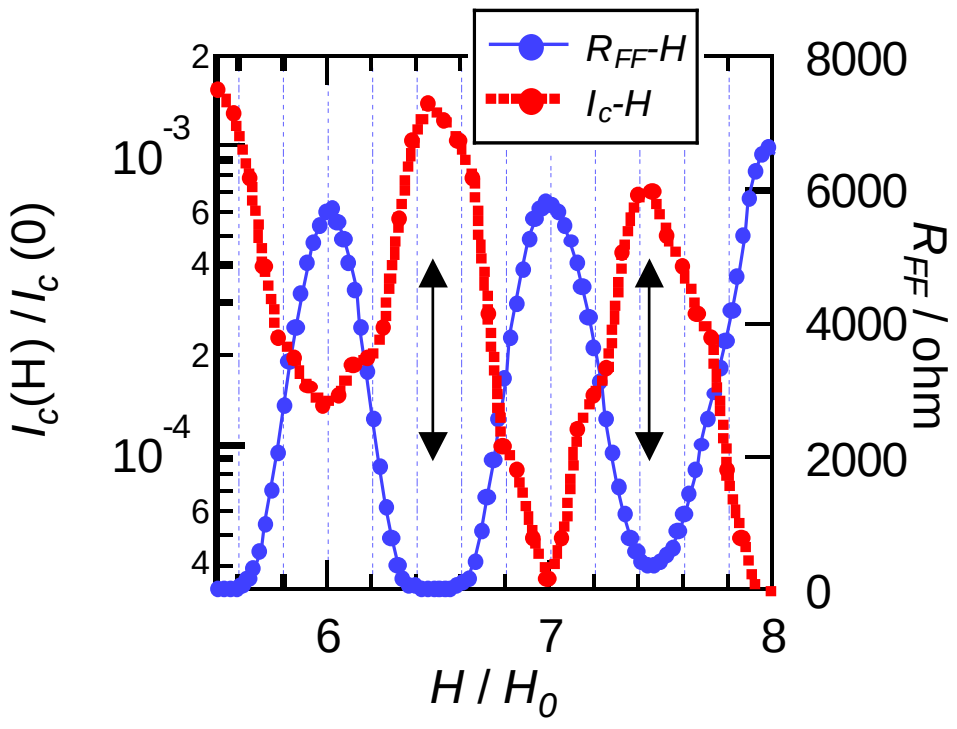

Fig. 2: Magnetic field dependence of critical current Ic (dotted-line) and flux flow resistance RFF (solid-line) of IJJs of L=1.8 mm at 10 K. The peaks or valleys in the H0 oscillations appear at the valleys or peaks in the Ic-H curve showing Fraunhofer dependence.

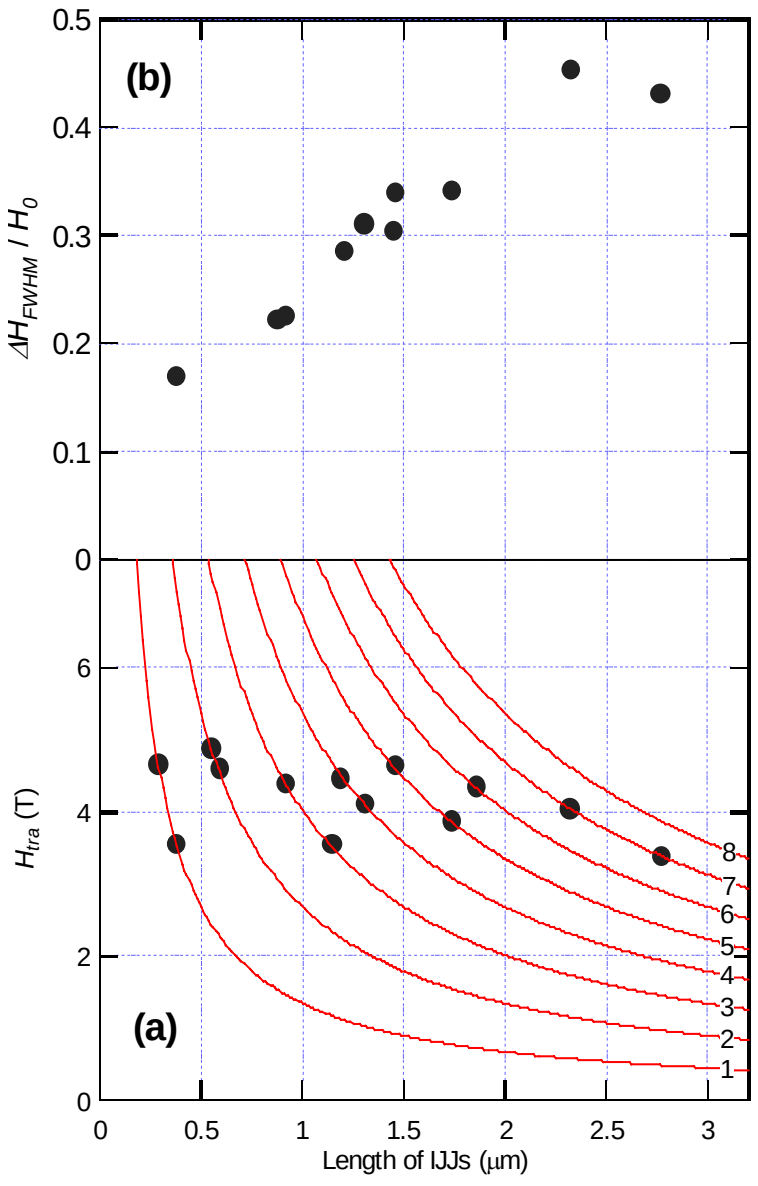

Fig. 3 (a): The length dependence of transition fields Htr marked by the arrows in Fig. 1. The hyperbola lines shows the number of Josephson vortices per junction. (b): The length dependence of full width at half maximum DHFWHM of the arrowed peaks indicated in Fig. 1.

of JV ($\gamma s$) overlaps, which occures above $H_1 \equiv \frac{\Phi_0}{\gamma\cdot s^2}$ as shown in Fig. 4 (a) (Here, $\gamma s \times s$ provides the area where non-linear core fully occupies each junction). In IJJs, JVLs generally form triangular JVLs due to the layered structure unlike in the case of SJJ. Above $H_1^* \equiv \frac{\sqrt{3}}{2}\frac{\Phi_0}{\gamma\cdot s^2}$, $J_{intra}$ becomes stronger than $J_{inter}$ considering the distance between the center of JVs as shown in Fig. 4 (b) (here, $J_{intra}$ and $J_{inter}$ expresses intra-layer and inter-layer interaction of JVs. $H_1$* is the field where JVL forms equilateral triangle lattice considering $\gamma$.) [12, 16]. The dominant $J_{intra}$ indicates that the interaction becomes one-dimensional in the $y$-direction (SJJ-like). When field reaches the $H_1$ where non-linear core starts to overlap, much strong $J_{intra}$ is expected. Considering $\lambda_c$, almost uniform magnetic field penetrates into our samples. Under the situation, phase difference between the neighbor superconducting layers changes linearly along the $b$-direction, which results in sinusoidal Josephson current distribution in each junction (This indicates short junction behavior known for SJJ). Therefore, above $H_1$, IJJs behave like individual SJJs in the short junction limit, which results in the appearance of Fraunhofer pattern. Due to the inverse relation to the Fraunhofer pattern, $R_{FF}$ shows peak at n$H_0$ (above $H_1$) because of minimal edge Josephson current, that is, $H_0$ oscillations appear. Therefore, for the appearance of peak at $n\cdot H_0$ as $H_0$ oscillations, $H_1$ should be satisfied. This indicates that $H_1$ becomes transition field above which $H_0$ oscillations appear. When we calculate $H_1$ using $\gamma$~200, $H_1$ is around 4.5 T, which agrees with the experimental data of $4.2\pm0.5$ T. So we can conclude that $H_0$ oscillations originate from the periodic suppression of $R_{FF}$ by the edge current arising from the Fraunhofer pattern.

In addition to the above-mentioned size-independent feature, we have found significantly size-dependent feature; The full width at half maxima of $H_0$ oscillations ( $\Delta H_{FWHM}$ ): As clearly seen in Fig. 1, the field modulation of $R_{FF}$ is remarkably dependent on the junction length. For a quantitative comparison, we analyzed $\Delta H_{FWHM}$ at the $H_{tr}$ (see Fig.1). As presented in Fig. 3(b), $\Delta H_{FWHM} / H_0$ becomes smaller with decrease of the $L$, and the field ranges at valleys tend to be larger accordingly. This is because the ratio of JVs which are directly affected by edges to the total number of JVs increases with the decrease of $L$, making, the edge pinning relatively stronger. Machida predicts by the numerical simulations that: (1) the triangular JVLs appear at $nH_0$ and rectangular JVLs at $(n+1/2)\ H_0$, and (2) rectangular JVLs become major structures in sub-micron IJJs [10]. This is consistent with the size-dependence of $\Delta H_{FWHM} / H_0$ in our experimental observation.

## IV. Conclusion

We have fabricated IJJs in Bi2212 single crystal whiskers with various lateral sizes down to the size of non-linear core of the JVs (0.29-2.77 μm), and investigated their periodic oscillations of flux flow resistance. We have found a size-independent critical field $H_1 = \frac{\Phi_0}{\gamma \cdot s^2}$ . Above the $H_1$, $I_c$ follows Fraunhofer dependence in narrow-length IJJs studied here, and then $H_0$ oscillations appear. In addition, the periodic oscillations of flux flow resistance ($R_{FF}$) become pulse-like in sub-micron IJJs. This indicates that rectangular JVL becomes major structures because of enhanced edge effects in sub-micron IJJs. From the experimental results obtained in this study, we expect that sub-micron IJJs have potential for devices with THz waveband application because of their stability of rectangular (in-phase) JVL structures.

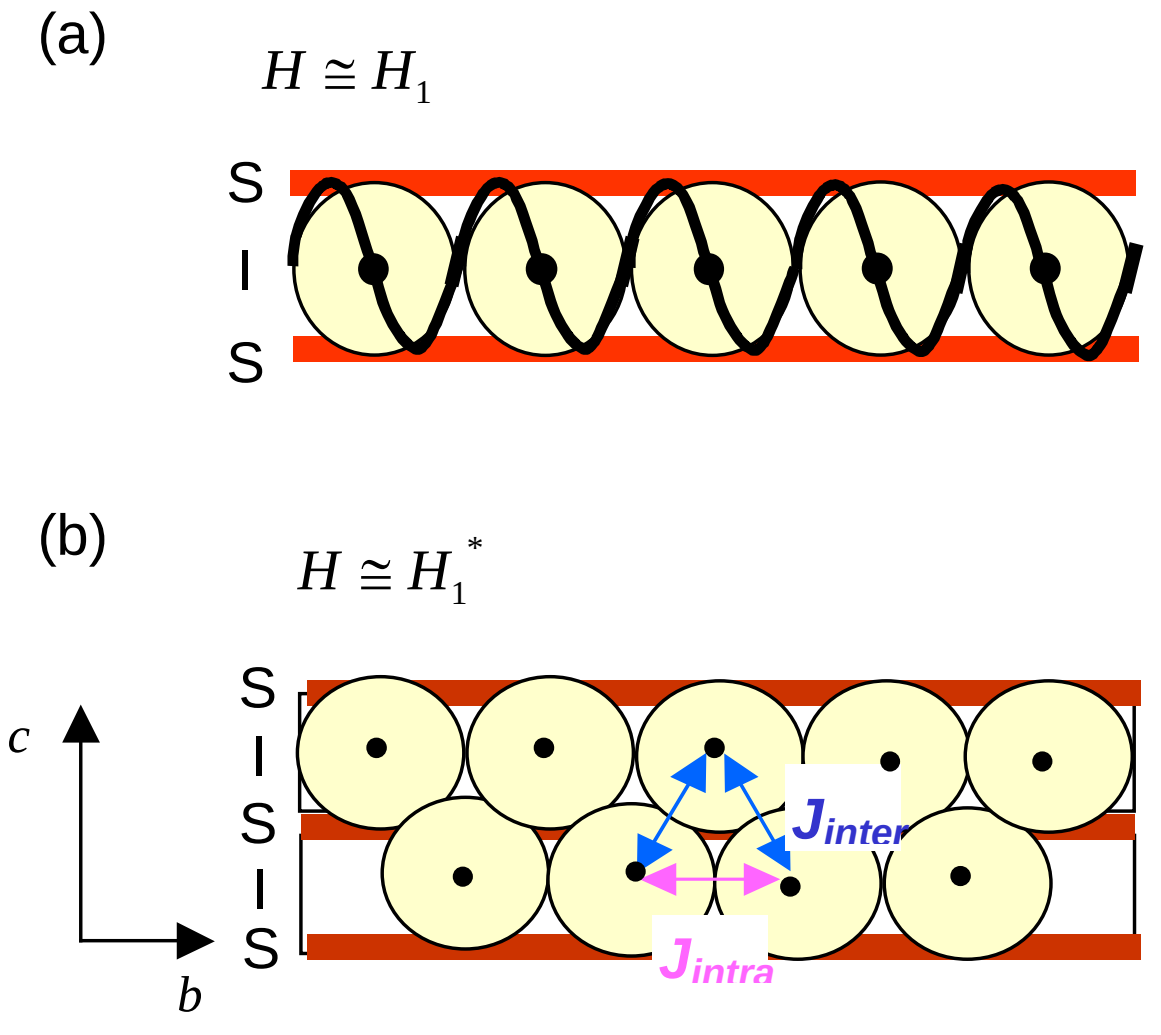

Fig. 4 (a): Schematic view of Josephson current distribution inside a single Josephson junction at . The black circle shows the size of the non-linear core of the Jopsephson vortex. S and I mean superconducting layer and insulating layer, respectively. (b): Schematic view of Josephson vortex lattice inside of IJJs at (projection to the y-gz plane where g expresses the anisotropy factor). The arrows express interaction between JVs.

## Acknowledgment

The authors would like to thank Dr. M. Machida of the Japan Atomic Energy Research Institute and Profs. I. Iguchi, M. Tachiki, T. Yamashita and Drs. X. Hu, Y. Takano of the NIMS and Prof. F. Nori of RIKEN for useful discussions, and Ms. R. Yamashita of the NIMS for critical reading of the manuscript.